\documentstyle[12pt, psfig, aasms4]{article}

\begin{document}

\title{
Gamma-ray burst: probe of a black hole}

\author{Wei Wang and Yongheng Zhao}
\affil{National Astronomical Observatories,
Chinese Academy of Sciences, Beijing, 100012}
\affil{wwang@lamost.bao.ac.cn}

\begin{abstract}
There is strong evidence for the existence of black holes(BHs) in some
X-ray binaries and most galatic nuclei, based on different measuremental
approaches, but black holes aren't finally identified for the lack of very
firm observational evidence up to now. Because the direct evidence for BHs
should come from determination of strong gravitational redshift, we hope
an object can fall into the region near the BH horizon where radiation can be
detected. Therefore the object must be compact stars as neutron stars(NSs),
then the intense astrophysical processes will release very high energy
radiation which is transient, fast-variant. These phenomena may point to
gamma-ray bursts(GRBs) observed. And recent observations of iron lines
suggest that afterglows of GRBs show the similar property in active galatic
nuclei(AGNs), implying GRBs may originate from intense events related to black
holes. In this letter, a model for GRBs and
afterglows is proposed to obtain the range of gravitational redshifts($z_g$)
of GRBs with known cosmological redshifts. Hence, we provide a new method that
with a search for high-energy emission lines(X- or $\gamma$-rays) in GRBs, one
can determine the gravitational redshift. We expect $z_g>0.5$ or even larger,
so that we can rule out the possibility of other compact objects such as NSs,
suggesting that the central progenitors of GRBs should be black holes.
\end{abstract}


\section{Introduction}

The approaches to search for observational evidence for the existence of black
holes are provided with the various measuremental methods, such as stellar
dynamics(Ghez et al. 1998, 2000), optical emission lines from gas
disks(Ferrarese, Ford \& Jaffe 1996; Macchetto et al. 1997), water maser
disks(Miyoshi et al. 1995), X-ray lines(e.g. Fe K$\alpha$ line, Nandra et al.
1997) and the strength of ultrasoft component of X-ray spectra(Zhang, Cui \&
Chen 1997). However, all these methods are only related to gaseous processes
where(at least two gravitational radius) the gravitaional field is weak. To
probe the very strong gravitational field, the radiation should result from
the process near the BH horizon where a large number high-energy photons can
be released. If the high-energy emission lines are identified, we can
determine the large gravitaional redshift. Here, we propose that GRBs may be
the best approach. 

Our method is partly motivated by the recent observational evidence for the
existence of Fe K$\alpha$ lines in the X-ray afterglows of GRB
990705(Amati et al. 2000), 990712(Frontera et al. 2001), 991216(Piro et
al.2000) and 000214(Antonelli et al. 2000), and in fact, GRB 970508(Piro et
al. 1999) and 970828(Yoshida et al. 1999) have also shown the evidence for the
iron lines. According to the observed line fluence, the mass of iron should
account for about $10^{-4}-0.1M_\odot$, where $M_\odot$ is the mass of the
sun,  which contradicts with the standard fireball model for GRBs because no
enough irons in the interstellar medium can be around the progenitors and if
the irons are produced during the bursts, it will involve the famous problem
of bayon contamination. To explain the iron lines, some researches have
proposed some mechanisms: with the energy injection, only a small mass of Fe is
required(Rees \& M\'esz\'aros 2000) and Fe comes from supernova which requires
that supernova explosion precedes the GRB event by several months to years
(Antonelli et al 2001).

However, the existence of iron lines shows us the some
similarity of GRB afterglows to AGNs, implying that the origin of GRBs and
their afterglows may be related to black holes and accretion process.
In this letter, we propose a model of GRBs, in which
the massive black hole captures a neutron star to produce a GRB and a normal
star forming an accretion disk to produce the afterglow. Then an alternate
mechenism is also proposed to explain the iron line origin: Fe line may come from
the disk formed by the normal star.

In the following section, we will give a brief discription of our GRB and
afterglow model where we will show that some GRBs may be the good candidates
of black holes. In Section 3, the method to probe the firm evidence for black
holes is provided, summary and outlook are presented in Section 4.

\section{GRB and Afterglow Model}

Our GRB model is simply described as follows. A massive black hole
catches a neutron star, which can produce a large number of $\gamma$-ray
photons through the intense astrophysical process in a short timescale. In this
case, we could observed only the burst. And the minimum variability
timescales of burst light curves are related to the NS size, $\Delta t
\geq {R\over c}(1+z)(1+z_g)$, of the order of milliseconds which is
consistent with GRB observations, where $R$ is the NS radius, $c$ is light
speed and $z$ is the cosmological redshift. If a binary (a neutron star and a
normal star) rather than an isolated neutron star falls into a black hole, when
the normal star is disrupted forming a disk and accreted into the black hole,
X-ray and optical radiation are released. Then both bursts and afterglows are
observed. Of course, only normal stars captured by black holes in
galactic centre will produce flares in optical or X-ray band without bursts,
which also may be of observability (Rees 1988, 1990). And our group now is
trying to survey those optical flares.

The afterglows will be very similar to the continuum radiation of
AGNs, for example, the spectrum of the afterglow of GRB 970508 (Galama, T.J.
et al. 1998a) may like that of a blazar. What's more, the accretion process can
account for the afterglows decaying in time as a power-law. After the tidal
disruption of a normal star occurs, two processes will supply mass to the
central black hole. The first has been studied by Rees (1988, 1990): the
stream of stellar mass strung out in far-ranging orbits, showing that the
infall rate declines as $t^{-5/3}$. The other case involves mass loss from teh
inner edge of the accretion disk. Cannizzo et al.(1990) have discussed
this case of late-time evolution of $\dot{M}$ in which the accretion disk
supply rate varies as $t^{-1.2}$.

BH-NS interaction has been studied by some previous authors with the
numerical simulations (Lattimer \& Schramm 1976; Klu\'zniak \& Lee 1998; Janka
et al. 1999), which was also taken as one of successful models of GRBs. Here,
we also have analyzed the process with simple calculations. Assuming the
different masses of black holes, we derived the gravitational redshifts
by computing the gravitational radius $r_g$ ($={2GM\over c^2}$, where $G$ is
gravitational constant) and the critical radius $r$ where a neutron star is
disrupted by the central black hole through tidal force. In calculations, we
suppose the mass and radius of the neutron star are 1.4 $M_\odot$ and 10 km
respectively. For a comparison,  the process is computed with  Landau
potential (Landau \& Lifshitz 1975):  
\begin{equation}
\phi=-{c^2\over2}{\rm ln}(1-{r_g\over r})
\end{equation}
and pseudo-Newtonian potential (Paczy\'nski \& Wiita 1980):
\begin{equation}
\phi=-{GM\over r-r_g}
\end{equation}
in separation. In Figure 1, two dashed lines have shown $z_g$ as a function of
the BH masses according to Landau(up) and pseudo-Newtonian(down) potential
respectively. We note $z_g \gg 1$ when masses are very large, implying that
massive black holes may be the better test for observations. 

From the above analysis, a black hole capturing an isolated neutron stars
produces a large number of high-energy photons through a very intense process,
which is a GRB observed. With the known total energy of
GRBs, we evaluated the gravitational redshift around the critical radius
\begin{equation}
1+z_g \leq {E_{\rm NS}\over E_\gamma}, 
\end{equation}
where $E_{\rm NS}$ is the total rest energy of a neutron star and $E_\gamma$ is
the isotropic $\gamma$-ray energy.  
Since $1+z_g = (1-{r_g\over r})^{-1/2}$, we can find 
$r-r_g \sim {r_g\over(1+z_g)^2-1}$. Hence, we estimate the
duration of a burst,  
\begin{equation}
T_\gamma \geq {r-r_g \over c}(1+z_g) \sim {M \over 10^5M_\odot}
{1+z_g \over (1+z_g)^2-1}, 
\end{equation} 
where $M$ is the mass of the black hole, which also tells us the shorter
bursts with the smaller central masses.  
The Equation(1) can give a low limit of $z_g$:
\begin{equation}
1+z_g-(1+z_g)^{-1} \geq {1\over T_\gamma}{M\over10^5M_\odot}. 
\end{equation}
Then the determination of $z_g$ will depend on our constraint on the masses of
black holes. With Eqs. 3 and 5, we can also
give a constraint on the masses of black holes,
\begin{equation} 
{M\over M_\odot} \leq 10^5
T_\gamma({E_{\rm NS}\over E_\gamma}-{E_\gamma\over E_{\rm NS}}).
\end{equation}

When a binary is falling into a black hole, the normal star will be disrupted
to form a accretion disk in the zone far away from the black hole which is very
similar to the accretion disks in AGNs, and afterglows in X-ray, optical
and radio bands can be produced in the disk from where Fe K$\alpha$ lines are
also emitted. We consider a thick disk model, then the radial velocity $v_r
\sim \alpha c_s$, where $\alpha$ is the viscosity parameter ($0 < \alpha \leq
1$) and $c_s$ is the sound speed. The viscosity timescale is given by
$t_{\rm vis} \sim {r_*\over v_r} \sim {r_*\over\alpha c_s}$, where $r_*$ is the
disk radius. $c_s$ can be estimated from $c_s^2 \sim {kT\over m_p}$, where $k$
is the Boltzmann constant, $T$ is the disk temperature and $m_p$ is the mass
of the proton. Since the radiation of afterglows is emitted from the disk, one
can find (Frank, King \& Raine 1992)  
$T \geq ({L\over 4\pi r_*^2 \sigma})^{1/4}$, where $L$ is the disk luminosity
also the afterglow luminosity, and $\sigma$ is the Stefan-Boltzmann constant.
Therefore, we reach an important result on the masses of black holes: 
\begin{equation}
{M \over M_\odot} \geq ({L \over 10^{23}{\rm erg s^{-1}}})^{1/10}
t_{vis}^{4/5} \equiv M_d, 
\end{equation}
where we have taken $\alpha \sim 1$, and $r_* \sim 3r_g$.

To further test our model, we firstly calculated the isotropic $\gamma$-ray
energy ($E_\gamma$) and optical (R band) peak luminosity ($L$) of 20
well-studied GRBs with known cosmological redshifts (Bloom, Kulkarni \& Djorgovski 2000), 
and derived some intinsic parameters of GRBs which are dispalyed
in Table 1. In the calculation, we accept the cosmological model with
$H_0=65{\rm km\ s^{-1} Mpc^{-1}}, \Omega_M = 0.3, \Omega_\Lambda = 0.7$,
taking $t_{vis} \sim t$, the timescale when the lightcurves of afterglows
evolve from the peak flux to the half. With these parameters and
above equations, we can estimate the ranges of the black hole masses and
$z_g$ in our model. And R band lumilosity also gives a
check on BH masses using the Eddington luminosity limit:
\begin{equation}
{M_L\over M_\odot} \sim {L\over 1.3\times10^{38}{\rm erg s^{-1}}}. 
\end{equation}
Our final results are presented in Table 1 and plotted in Figure 1, in which
two dashed lines have been noted above. From Figure 1, we may draw following
three conclusions. First, we have found gravitational redshifts of some GRBs
are very large: seven GRBs whose gravitaional redshifts $z_g > 0.5$, which
strongely implies that the central engines of these GRBs should be the best
candidates of black holes. Second, the masses of black holes mostly distribute
around $10^6 M_\odot$ which are comparable to the masses of black holes in the
normal galaxies as in our Galaxy. Third, the range of $z_g$ of GRBs calculated
is consistent with the range given by two lines, supporting our model.

Because short bursts (with the duration $T_\gamma<2{\rm s}$) correspnd to the
relatively small masses of black holes, radiation from the disk around
low-mass black holes will be so faint that it cannot be observed by our
optical telescopes. Therefore, our model for GRBs and
afterglows may give a reliable explanation why afterglows cannot be observed
from most detected GRBs but only in a very small part with the
longer durations. Because in galaxies the number of isolated neutron stars is
larger than the number of neutron stars in binary systems according to the
observations of pulsars (Taylor, Manchester \& Lyne 1993), the possibility of
the former captured by black holes will be much larger, implying the
afterglows of GRBs are really rare, which is consistent with the
observations (Lamb 2000). The different actions between long-duration GRBs and
AGNs may result from the circumstance of black holes, since AGNs are in a
dirty environment while GRBs in a clean one with a neutron star
or binary captured occassionally. We think that our unified model as a possible mechanism of
GRBs and afterglows can successfully explain the different occuring rates of
bursts and afterglows and the formation of iron lines recently discovered.

\section{Searching for Firm Evidence for Black Holes}

In the above section, we have applied our model to some GRBs
with known cosmological redshifts. In Table 1, we note that some GRBs have
very large gravitational redshifts such as GRB 970828 ($z_g>3.0$),
980425 ($z_g>1.4$), 000301c ($z_g>1.9$). Therefore, in our model, these GRB
central engines should be the good candidates of black holes. However, to
further probe the central body of a gamma-ray burst, we expect that high
spectral resolution detectors can find the reliable emission lines in the
spectra of GRBs and also suggest that pair annihilation lines (511 keV line)
may be the best approach. Hitherto, the emission lines have been detected and
identified in the energy spectrum of GRBs by Venera 11 and 12 (Mazets et al.
1981) before 1980 and BASTE (Briggs et al. 1997, 1999) in 1990's. The centroid
energies of emission lines distribute in two different ranges, 330-460 and
40-50 keV.

The lines in the broad range of 330-460 keV were interpreted as the
strongly redshifted 511 keV annihilation lines, but the emission lines at 40-50
keV were thought to be formed in another process. For instance,  
GRB 790526 revealed a emission line at 45 keV which was previously
thought to be the reliable evidence of emissions at the cycloton frequence in
the surface of a nearby magnetized neutron star (Mazets et al. 1981). 
However, since the measurement of the redshift of GRB 970508 (Metzger et al.
1997), the evidence for a cosmological distance scale for most or all bursts is
comfirmed, then  GRBs cannot be a local phenomina on the surface of a neutron
star. Hence, we need a new physical mechanism to explain the emission lines
observed. In this letter, we interprete the emission lines as strongly
redshifted 511 keV annihilation lines in the gravitational field of black
holes. Supposing the cosmological redshift of GRBs, $z \sim 1$, we calculated
the gravitational redshifts of the lines whose centroids are around 50 keV,
$z_g \sim 4$ in accordance with our model expection. Because the gravitational
redshift produced by the known compact objects including neutron stars or even
possible strange stars cannot be over $0.5$,  we may make our decision
that the central bodies of these bursts are the best candidates of black
holes. In fact, the other lines at 330-460 keV observed have the gravitational
reshifts in the range 0.1-0.5, which is also very difficultly realized on the
surface of a neutron star. Therefore, GRBs may really originate in the region
with the very strong gravitational field, where the central progenitors are
black holes with the captured neutron stars.

\section{Discussion and Summary}

Because the iron lines discovered in
the X-ray afterglows of some bursts imply the relation between AGNs and GRBs,
a possible model of GRBs and afterglows is proposed in the letter.
In our model, we can determine the range of gravitational redshifts of
$\gamma$-ray radiation with the intrinsic parameters of GRBs and afterglows,
and we note some of them are very large ($z_g>0.5$) implying GRBs as
probes of ultra-strong gravity of black holes. However, very firm evidence
should come from the identification of GRB line emissions. Through the
determination of the very strong gravitational redshift of emission lines, one
can identify the central bodies as black holes.
Thus, the spectral line observation of GRBs can provide a very powerful
evidence for the existence of black holes. Because of the level limit on the
present missions, we need more advanced detectors with higher spectral
resolutions in future. Here, we hope that recently launched HETE-II and
Swift(in 2003) will do much work on the search for GRB emission lines

The model of GRB afterglows in this letter is also used to explain the X-ray flares recently detected in several nearby normal galaxies with the ROSAT data base. Assuming the number of massive stars which can produce NSs through supernova explosion is about $10\%$ of total star number and one supernova event occurs in 100 years, we can simply estimate the density ratio of stars and NSs in a normal galaxy as $n_{\rm star}/n_{NS}\sim 1000$. The observations have given the GRB rate $R_{\rm GRB}\sim 10^{-6}-10^{-7}{\rm /yr/gal}$, so we can approximately obtain the flare event rate $R_{\rm flare}\sim 10^{-3}-10^{-4}{\rm /yr/gal}$ which is consistant with the theoretical expected tidal disruption rate (Rees 1988, 1990) and the searching results given by ROSAT all-sky survey (Komossa \& Dahlem 2001). With the flare event rate and assuming a solar mass is swallowed by the black hole per event, we find that the central black hole masses in normal galaxies can increase at least up to $10^6-10^7{\rm M_\odot}$ (such as the massive black holes in the center of our Galaxy and M31) through the tidal disruption of stars within the Hubble time scale ($\sim 10^{10}$yr). This conclusion is also important to the formation and evolution of normal galaxies and massive black holes.

{\bf Acknowledgments}
We would be grateful to Mr.Ye Fangfu for the help in the calculations. This
research is supported by the National Natural Science Foundation of China.

\pagebreak

\scriptsize

\begin{figure}
\centerline{\psfig{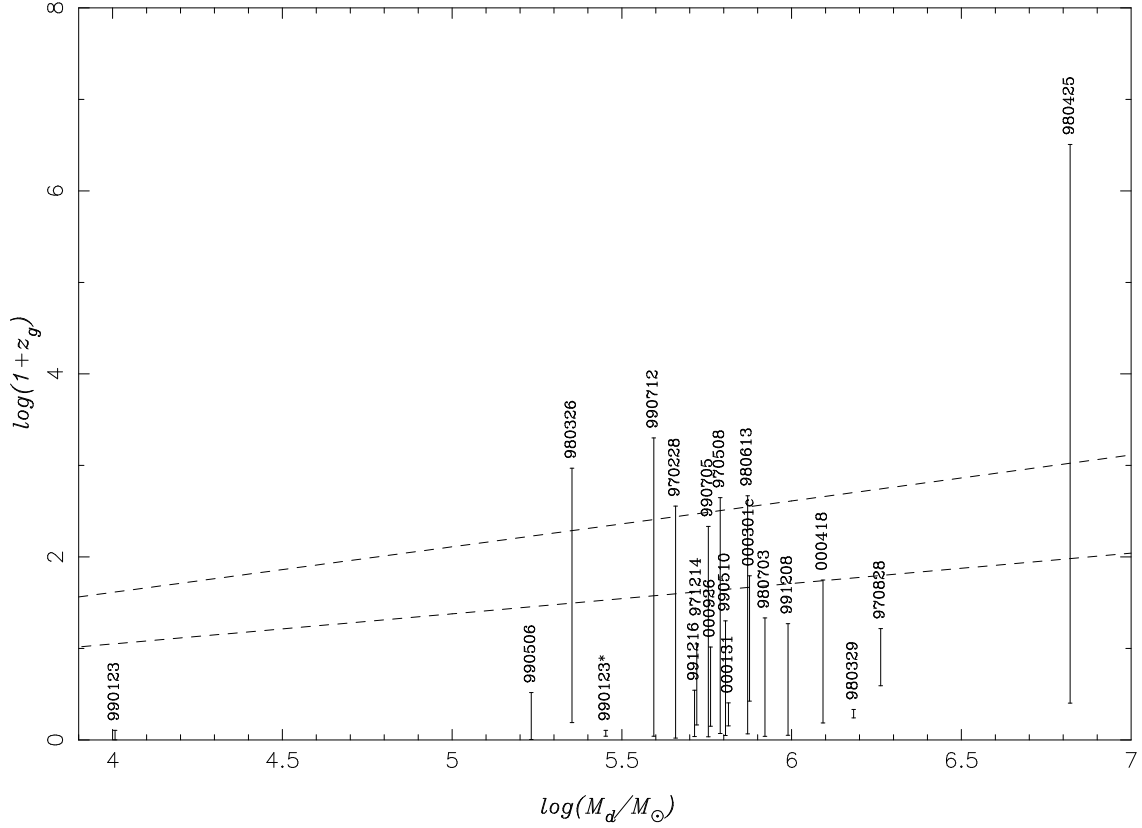}}
\caption[]{\protect The range of GRB gravitational redshifts estimated from our
model are plotted as a function of the lower limits of the BH masses. 
Two dashed lines describe the critical gravitational redshift of
neutron stars disrupted by tidal force according to Landau(up) and
pseudo-Newtonian(down) potential in separation. 20 GRBs with known
cosmological redshifts are displayed, and we noticed most masses distribute
around $10^6M_\odot$ except GRB 980425 and 990123(as noted in Table 1). If we
cancel the optical flash and only take the second afterglow of 990123, we note
that 990123$^*$ which is also plotted in the figure become a normal one
similar to others. Refer to the text for details.}   
\end{figure}

\pagebreak

\begin{deluxetable}{lccccccc}
\tablecaption{The intrinsic parameters of 20 GRBs with known cosmological 
redshifts and derived ranges of black hole masses and
gravitational redshifts determined in our model.}
\tablehead{
\colhead{name} & \colhead{$T_\gamma$(s)} & \colhead{$log E_\gamma$(erg)} & 
\colhead{$log t(s)$} & \colhead{$log L$(erg\ s$^{-1}$)} & \colhead{$log
M$($M_\odot$)} & \colhead{$z_g$} & \colhead{$z$}} 
\startdata
970228 & 47 & 51.9 & 4.5 & 43.2 & 5.7-9.2(5.1) & 0.05-358 & 0.695 \\ 
970508 & 18.7 & 51.8 & 4.5 & 45.1 & 5.9-8.9(7.0) & 0.16-443 & 0.875 \\ 
970828 & 5 & 53.2 & 5.0 & 45.6 & 6.3-6.9(7.4) & 3.0-15.5 & 0.935 \\ 
971214 & 6.8 & 53.4 & 4.3 & 45.8 & 5.8-6.9(7.7) & 0.6-10.2 & 3.42 \\ 
980326 & 2.5 & 51.5 & 4.0 & 44.5 & 5.4-8.3(6.4) & 0.6-1000 & 1 \\
980329 & 13 & 54.2 & 4.9 & 45.6 & 6.2-6.3(7.4) & 0.8-1.1 & 3.5 \\
980425 & 31 & 47.9 & 6.0 & 45.2 & 6.8-13.0(5.1) & 1.4-3$\times 10^6$ & 0.0085 \\
980613 & 24 & 51.8 & 4.7 & 44.1 & 5.9-9.1(6.0) & 0.16-466 & 1.096 \\
980703 & 46 & 53.1 & 4.7 & 44.6 & 6.0-8.0(6.4) & 0.1-20.5 & 0.966 \\
990123 & 15 & 54.3 & 1.6 & 50.2 & 4.1-6.0(12) & 0.01-0.3 &1.61 \\
990123$^*$ & \ & \ & 3.5 & 46.3 & 5.4-6.0(8.0) & 0.15-0.3 & \ \\ 
990506 & 65 & 53.9 & 3.7 & 45.7 & 5.7-7.3(7.6) & 0.05-2.3 & 1.310 \\
990510 & 28.6 & 53.1 & 4.3 & 46.6 & 5.6-7.5(7.5) & 0.1-19 & 1.62 \\
990705 & 36 & 52.1 & 4.7 & 43.0 & 5.7-8.9(4.8) & 0.1-214 & 0.25 \\ 
990712 & 21 & 51.1 & 4.3 & 45.7 & 5.6-9.6(6.4) & 0.1-2000 & 0.434 \\
991208 & 41 & 53.2 & 4.7 & 45.3 & 6.0-7.9(7.2) & 0.15-17.7 & 0.7055 \\
991216 & 30 & 53.9 & 4.3 & 45.7 & 5.7-7.0(7.6) & 0.1-2.5 & 1.02 \\
000131 & 9 & 54.1 & 4.5 & 45.3 & 5.8-6.2(7.2) & 0.4-1.5 & 4.5 \\
000301c & 3.3 & 52.6 & 4.5 & 45.9 & 5.9-7.3(7.8) & 1.9-61 & 2.0 \\
000418 & 14 & 52.7 & 4.9 & 44.7 & 6.1-7.9(7.6) & 0.6-55 & 1.11854 \\ 
000926 & 8.2 & 53.4 & 4.3 & 46.2 & 5.8-7.0(8.1) & 0.4-9 & 2.066 

\enddata

\tablecomments{The timescales $T_\gamma$ and $t$ have been devided by a
factor $(1+z)$ correction to the GRB intrinsic timescales. Since a
particular prompt optical flash (Akerlof et al. 1999) was observed 15 seconds
after GRB 990123 stated, we take two cases of the afterglow: the
flash($m_R\sim9$) as a peak and the second afterglow($m_R\sim17$) as the peak
where $m_R$ is is the visual magnitude of optical peak in R band. Because of
the possible association between GRB 980425 and SN1998sw (Galama et al. 1998b),
$z$ is very small making the $\gamma$-ray energy too low, and $t$ is very long
probably due to the effect of the light curve of SN1998sw. $M_L$ is shown in
the brackets, in many cases, it shows a super-Eddington accretion, but
some are the sub-Eddington ones.}
\end{deluxetable}


\begin{references}

\reference{}
Akerlof, K, et al., 1999, Nature, 398, 400 

\reference{}
Amati, L. et al., 2000, Science, 290, 953 

\reference{}
Antonelli, L.A., et al., 2000, ApJ, 545, L39 

\reference{}
Bloom, J.S., Kulkarni, S.R. and Djorgovski, S.G., 2000, preprint(astro-ph/0010176)

\reference{}
Briggs, M.S., et al., 1997, preprint(astro-ph/9712096)

\reference{}
Briggs, M.S. et al., 1999, preprint(astro-ph/9901224)

\reference{}
Cannizzo, J.K., Lee, H.M. and Goodman, J., 1990, ApJ, 351, 38 

\reference{}
Ferrarese, L., Ford, H. C., and Jaffe, W.,1996, ApJ,  
470, 444 

\reference{}
Frank, J., King, A.R. and Raine, D.J., 1992, {\it Accretion Power in
Astrophysics, Second Edition}, p.5

\reference{}
Frontera, F., et al., 2001, preprint(astro-ph/0102234)  

\reference{}
Ghez, A.M. et al., 1998, ApJ, 509, 678 

\reference{}
Ghez, A. M., et al., 2000, Nature, 407, 349 

\reference{}
Galama, T.J., et al., 1998a, ApJ, 500, L97

\reference{}
Galama, T.J., et al., 1998b, Nature, 395, 670 

\reference{}
Janka, H.-T., Eberl, T., Ruffert, M. and Fryer, C.L., 1999, 
preprint(astro-ph/9908290)

\reference{}
Kamossa, S. and Dahlem, J., in {\it MAXI workshop on AGN variability}(astro-ph/0106422)

\reference{}
Klu\'zniak, W. and Lee, W. H., 1998, ApJ, 494, L53

\reference{}
Lamb, D.Q., 2000, preprint(astro-ph/0005028) 

\reference{}
Landau, L.D. and Lifshitz, E.M., 1975, {\it The Classical Theory of
Fields}, p.253

\reference{}
Lattimer, J. M. and Schramm, D. N., 1976, ApJ, 210, 549 

\reference{}
Macchetto, F., et al., 1997, ApJ, 489, 579 

\reference{}
Mazets, E.P. et al., 1981, Nature, 290, 378

\reference{}
Metzger, R., et al., 1997, Nature, 387, 878 

\reference{}
Miyoshi, M., et al., 1995, Nature, 373, 127 

\reference{}
Nandra, K., et al., 1997, ApJ, 477, 602 

\reference{}
Paczy\'nski, B. and Wiita, P.J., 1980, A\&A, 88, 23 

\reference{}
Piro, L., et al., 1999, ApJ, 514, L73 

\reference{}
Piro, L., et al., 2000, Science, 290, 955

\reference{}
Rees, M.J., 1988, Nature, 333, 523

\reference{}
Rees, M.J., 1990, Science, 247, 817 

\reference{}
Rees, M.J. and M\'esz\'aros, P., 2000, ApJ, 545, L73  

\reference{}
Taylor, J.H., Manchester, R.N. and Lyne, A.G., 1993,
ApJS, 88, 529

\reference{}
Yoshida, A. et al., 1999, A\&AS, 138, 433

\reference{}
Zhang, S.N., Cui, W. and Chen, W., 1997, ApJ, 482, L155 


\end{references}
\end{document}